\documentclass[12pt,preprint]{aastex}

\newcommand\lam{\mbox{$\:\lambda $ }}
\newcommand\lamlam{\mbox{$\:\lambda\lambda $ }}
\newcommand\ha{{H$\alpha$}}

\newcommand\al{$\alpha$}

\newcommand\kms{\:\rm{\,km\,s^{-1}}}

\newcommand\masy{\:\rm{\,mas\:yr^{-1}}}

\newcommand\arcspix{\arcsec\:{\rm pixel}^{-1}}

\newcommand\ie{{\it i.e.}}

\newcommand\hi{\ion{H}{1}}

\newcommand\sii{[\ion{S}{2}]}

\newcommand\oiii{[\ion{O}{3}]}

\slugcomment{To appear in {\it The Astrophysical Journal}}

\shorttitle{Proper Motions in SNR G292.0+1.8 }
\shortauthors{Winkler et al.}

\begin{document}
\title{Expanding Ejecta in the Oxygen-Rich Supernova Remnant G292.0+1.8: Direct Measurement through Proper Motions}
\author{P. Frank Winkler\altaffilmark{1}}\affil{Department of Physics, Middlebury College, Middlebury, VT 05753}  
\email{winkler@middlebury.edu}
\author{Karl Twelker\altaffilmark{1} \altaffilmark{2}}
\affil{Department of Physics, Middlebury College, Middlebury, VT 05753}
\author{Claudine N. Reith\altaffilmark{1}}
\affil{Department of Physics, Middlebury College, Middlebury, VT 05753}
\and
\author{Knox S. Long\altaffilmark{1}}
\affil{Space Telescope Science Institute, 3700 San Martin Drive, Baltimore MD 21218}
\email{long@stsci.edu}

\altaffiltext{1}{Visiting Astronomer, Cerro Tololo Inter-American Observatory.
CTIO is operated by AURA, Inc.\ under contract to the National Science
Foundation.}
\altaffiltext{2}{Present address:  Department of Physics, Stanford University, Stanford, CA  94305}


\begin{abstract}
We report here the first study of proper motions of fast filaments in the young, oxygen-rich supernova remnant  G292.0+1.8, carried out using a series of  \oiii\  5007 \AA\  emission-line images taken over a period of more than 21 years.   Images taken at seven epochs from 1986 to 2008, all from telescopes at the Cerro Tololo Inter-American Observatory, show oxygen-emitting filaments, presumably ejecta fragments, throughout most of the remnant.  We have measured the proper motions for 67  discrete filaments through two-dimensional correlations between images from different epochs.   While the motions are small, mostly 20 to 100$\masy$, they are nevertheless measurable through a robust technique of averaging measurements from many epoch pairs.  The data are qualitatively consistent with a free-expansion model, and clearly show systematic motions outward from a point near the center of the radio/X-ray shell.  
Global fits using this model indicate an expansion center at 
${\rm R.A.}\:(2000.0) = 11^h 24^m 34.4^s,\  {\rm Dec.}\: (2000.0) = -59\degr 15\arcmin 51\arcsec$, and a kinematic age  of $2990\pm 60$\ years.
The young pulsar PSR J1124--5916 is located 46\arcsec\ southeast of the expansion center.  Assuming that it was launched by the supernova, we expect the pulsar to be moving southeastward at $16 \masy$, or a transverse velocity of $440 \kms$\@.
We find the fastest ejecta  along an axis oriented roughly N-S in the plane of the sky, suggesting that  a bipolar explosion produced G292.0+1.8, as appears to have been the case for Cas A.

\end{abstract}


\section{Introduction}
Massive stars produce copious quantities of oxygen and other low-Z elements during their hydrostatic evolution, and these are ejected when the stars end their lives in spectacular fashion as core-collapse supernovae (SNe).  After the pyrotechnics of the explosion fade, the ejecta coast outward at velocities of up to several thousand $\kms$, and the fastest of them interact with the surrounding circumstellar and/or interstellar medium and produce an outward propagating blast wave.  The interaction also results in the dramatic slowing of the ejecta and in a reverse shock that propagates inward, gradually encountering slower-moving ejecta and exciting them, until eventually the shock dissipates  as it nears the center.  As dense knots of ejecta encounter the reverse shock, they are excited and become visible, radiating emission lines characteristic of their composition---oxygen especially since this is the most abundant element in the outer core of typical progenitors.    The resulting oxygen-rich filaments---with a spectrum dominated by oxygen lines, showing little or no hydrogen, and with typical velocities $\gtrsim 1000 \kms$--- may be seen for a few hundred to a few thousand years.   Individual  filaments last for a much shorter time, as their material is evaporated through the interaction with the reverse shock to enrich the hot, X-ray-emitting plasma that fills most of the shells of these young supernova remnants (SNRs), producing an X-ray spectrum with strong lines from elements including Ne, Mg, Si, S, and others in addition to O\@.  Another product of a core-collapse SN is a compact remnant---a neutron star or perhaps a black hole---that may also be visible as a point, possibly pulsing, X-ray source, sometimes surrounded by a synchrotron-emitting pulsar-wind nebula.  Observationally, objects displaying these fast, O-rich optical filaments have become known as oxygen-rich SNRs, a small class including only Cas A, Puppis A, and G292.0+1.8 in the Galaxy plus  4 to 6 others in the Magellanic Clouds and other nearby galaxies.

Of these, the  only one to show all the above  characteristics expected from a core-collapse supernova is G292.0+1.8 (hereafter simply G292).  
G292 was first discovered and denoted as MSH11--5{\it4} in the radio survey by \citet{mills61}, and on the basis of its non-thermal spectrum was identified as a SNR by \citet{milne69} and \citet{shaver70}.   Much more recently, the most detailed radio images of G292 are by \citet{gaensler03}, who found  that the SNR has a diameter of 8\arcmin, and who also give what is probably the most reliable distance estimate: $6.2\pm0.9\;$kpc, based primarily on the \hi\ absorption profile.  Throughout this paper we will scale all distance-dependent quantities to $d_6\equiv d/(6\;{\rm kpc})$\@.   

X-rays from this source were first identified in HEAO-1 data by \citet{share78}, and its properties were investigated from the {\it Einstein} Observatory by \citet{clark80}, who found strong X-ray lines from Mg, Si, and S, and by \citet{tuohy82}, who noted a prominent central ring, possibly suggesting a rotating Type II SN precursor.   Most recently, studies of G292 from  {\it Chandra} have produced  spectacular 
high-resolution images  \citep{park02, park07}.  
The X-ray spectrum is dominated by K-shell lines of O, Ne, Mg, Si, and S.  Non-equilibrium 
ionization analyses require large enhancements (relative to solar) in abundances 
for O, Ne, and Mg, with lesser enhancements for Si, S, and Fe \citep{hughes94, gonzalez03}.
By comparing the inferred abundances with the integrated yields predicted by  models 
for core-collapse supernovae, these groups estimated a progenitor mass in the range $\sim25-40\:M_\sun$.
In a more detailed study of individual X-ray features,  \citet{park04} found that
different knots have very different  compositions and suggested that  these 
represent clumps of ejecta from different zones of the progenitor, and in a much deeper ACIS image \citet{park07} have found evidence for asymmetries in the distribution of (at least) the oxygen ejecta.  These latest images show that the  X-ray shell extends to the south, with a full extent of 9.6\arcmin\ (N-S)$\;\times\;$8.4\arcmin\ (E-W), or $16.7 \times 14.7\:d_6\;$pc. 

The original {\it Chandra} image of G292 also revealed a compact central source, 
surrounded by what appeared to be a pulsar wind nebula \citep{hughes01}.  
Shortly thereafter, \citet{camilo02} discovered the radio pulsar PSR J1124--5916 within
or very near G292\@.   \citet{hughes03} then showed that the compact X-ray source is pulsed with the same period, confirming that it must be the compact remnant of the star that produced G292.  
The period of 135 ms and spin-down age of 2900 yr are roughly consistent with the age of the G292 SNR, estimated by \citet{chevalier05} as 2700--3700 yr based on properties of the pulsar-wind nebula. 

\citet{goss79} discovered optical emission from G292, consisting solely of a group of filaments near the center of the radio shell  with a spectrum showing strong lines from O and Ne, and no lines from H, N, or S that are more typical of SNRs\@.   \citet{murdin79} showed that the brighter filaments have high radial velocities, $-650  \lesssim v_{rad} \lesssim +1380 \kms$, suggesting that they are undecelerated (or minimally decelerated) ejecta from the supernova explosion.  Together these results provided the first identification of G292  as an oxygen-rich SNR\@.   \citet{murdin79} estimated the remnant's  age as $\tau \lesssim D/\Delta v \sim 1800\:d_6\;$yr,  based simply on a scaling argument using a diameter $D\approx 2\arcmin \approx 3.5\:d_6\:$ pc, the extent of the central radio peak where they had observed the optical filaments, and their measured radial velocity range of $2030\kms$\@.  More limited spectra by \citet{vdb79} showed a velocity dispersion only slightly smaller, and further spectra by \citet{braun83} led to a similar kinematic estimate for the age, $\tau  \sim 2000\: d_6\;$yr.\footnote{Both \citet{murdin79} and \citet{braun83} used a distance of 5.4 kpc and hence found slightly smaller ages.  We have simply scaled their results to a uniform distance of 6 kpc.  Since they detected only the brightest optical filaments, near the center of the shell, their estimates for the diameter and scaled age are much smaller than more recent ones.}   

\citet{ghavamian05}, in an extensive kinematic study  using the Rutgers Fabry-Perot imaging spectrometer, reported \oiii \lam 5007 emission features from the central 7.5\arcmin\ of G292 throughout a velocity range of at least $-1440 \lesssim v_{rad} \lesssim 1700\kms$ (the full range covered in their etalon settings).   They found the  distribution of O-rich knots to be quite asymmetric, with virtually all of the knots north of the (radio-determined) shell center being blue-shifted, while toward the south there are both red- and blue-shifted knots.   From a plot of radial velocity vs projected radius, \citet{ghavamian05} found fair agreement with a model in which the knots are distributed around an expanding shell.   Fitting this distribution with a velocity-radius ellipsoid, they found an ejecta shell velocity $v_{ej} \approx 1700\kms$, centered about a systemic velocity of $v_c \approx +100\kms$.  Assuming a spherical shell geometry led to an estimated age of $\tau \equiv R_{ej}/V_{ej} = (3000 - 3400) d_6\:$yr, where the range reflects both the width of the velocity-radius ellipsoid and models using somewhat different assumptions.   The greater age, by almost a factor of 2 compared with \citet{murdin79} and \citet{braun83},  is now quite consistent with  the pulsar spin-down age and the estimate based on the PWN properties.  The change in the optically-based estimate is due almost entirely to the fact that \citet{ghavamian05} observed (fainter) O-rich filaments out to a radius roughly twice that found in the earlier studies.   

\notetoeditor{We refer several times to the Ghavamian et al. 2005 paper.  It may be desirable to use a shorthand reference for this paper, e.g. as GHW, or GHW05.  I will leave this to the discretion of the editors.}

The full optical extent of G292 was shown for the first time by \citet[][paper 1]{winkler06}, who found O-rich optical knots throughout most of the 8\arcmin\ shell seen in radio and X-ray images, with a few knots extending even beyond the the primary shell and into the southern extension seen in the recent deep {\it Chandra} image \citep{park07}.  In paper 1 we also found a few faint knots showing \sii \lamlam 6716, 6731 emission in addition to the oxygen lines, the first optical evidence for the presence of O-burning products in the ejecta.  Most of the outer knots exhibit radially-oriented pencil-like morphologies, suggesting an origin as Rayleigh-Taylor fingers.





In this paper we present our study of the proper motions of oxygen-rich knots in G292, the result of a series of CCD images from 1986 to 2008.  These clearly show the systematic  motion of the knots outward from a point near the center of the outer radio shell.   Global fits of a free-expansion model lead to what we believe to be the most reliable age measurement for the SNR, and indicate that the center is 46\arcsec\ from the pulsar PSR J1124--5916.   We discuss briefly the three-dimensional distribution of ejecta, which indicates a clear asymmetry and suggests that the progenitor explosion may have been a bipolar one.





%

\section{Observations and Data Reduction}
Emission-line images used in this study were taken in seven separate observing runs on various telescopes at the Cerro Tololo Inter-American Observatory (CTIO) from 1986 through 2008.  A journal of the observations is given in Table 1.  The runs from 1986 and 1991 used CCD chips with a small field of view that covered only the central region of G292, as noted in the table.  Those from 1999 onward all used the 0.9m telescope and Tektronix 2K\#3 chip, and covered a field 13.7\arcmin\ square, easily encompassing the entire remnant, at a scale of $0.401\arcspix$.   All the observations used narrow-band interference filters, with characteristics also listed in Table 1.  Note that while some filter nominally designed for \oiii \lam5007 was used in each case, we in fact used three different such filters, and even for the same filter the effective bandpass is different  when used in beams converging with different focal ratios.  In some cases, especially the runs on the 0.9m in 1999 and 2000, the filter was one designed for a beam faster than the 0.9m's of f/13.5; as a result some of the most blueshifted filaments are barely visible in the 1999 and 2000 images.   In addition to images in \oiii \lam5007, we also obtained images in a continuum band at each run from 1991 onward, so we could subtract away most of the myriad stars that pervade the crowded Galactic field of G292 and make it easier to measure the motions of the filaments.  

On the 1986  run we obtained only a pair of \oiii\ images, but  on all the subsequent runs we obtained three or more individual frames in both line and continuum filters, dithered by several arcsec between frames in order to remove bad columns and other systematic effects on combining the images.  All images from all epochs were processed using standard IRAF\footnote{IRAF is distributed by the National Optical Astronomy Observatory, which is operated by AURA, Inc., under cooperative agreement with the National Science Foundation} procedures for bias-subtraction and flat-fielding, the latter based on well-exposed dome or twilight sky flats.   During each set of observations we obtained several images of spectrophotometric standard stars from the list of  \citet{hamuy92}, and used these to flux-calibrate our images.

In order to place all the images on a common  world coordinate system (WCS), we used about 300 astrometric stars from the UCAC2 catalog \citep{zacharias03} (although many of these stars fell outside the fields of the earlier data).  The stellar positions were corrected for proper motions to the appropriate epoch using the software provided with the UCAC2 catalog.  We then calculated a precise WCS for each frame using the IRAF tasks {\it ccfind} and {\it ccmap} to find the centroid of the reference stars and then fit them to a common projection.  In the fitting process, we allowed for distortions from a standard tangent-plane projection of up to third order (using the ``tnx" projection in IRAF).  With the 0.9m frames we typically fit the positions of over 200 stars, with an rms dispersion of $\lesssim 50\;$mas in both directions.  For the earlier frames there were fewer reference stars, but equally good fits.

We then defined (arbitrarily) a standard  coordinate system:  a
tangent-plane projection centered at ${\rm R.A.}\,(2000.) = 11^h24^m31.0^s$, ${\rm Decl.}\,(2000.) = -59\degr 15\arcmin 30.0\arcsec$, with a scale of exactly $0.200\arcspix$, and each individual frame was transformed to the standard one using the IRAF task {\it wregister}\@.  All the transformations used bilinear interpolation and  double-precision arithmetic.  The new scale was chosen as significantly finer than the pixel size for any of our images, while maintaining images of manageable size.  At this point all the individual images from each run were combined to obtain a single image at each wavelength for each epoch.


Precise quantitative  measurement of the motions of knots is simplified greatly if we remove the stars, so that an individual knot can be isolated for the correlation analysis described in the next section.  For this we used the matched continuum images, also transformed to our standard system and combined in exactly the same way as the \oiii\ images.\footnote{For the 1986 image, where we had no matched continuum, we instead used a combination of continuum frames from later runs, suitably scaled and psf-matched.  While less than ideal, this nevertheless did a good job at removing most of the faint stars.}   The continuum image for each epoch was appropriately scaled and subtracted from the emission-line image, judging by eye what scaling factor did the best job of removing the stars.  If the stellar profiles of the emission-line and continuum image were not a close match, we used the IRAF task {\it psfmatch}, which determines a kernel to convolve with the sharper of the two images in order to match the profiles for stars in the less-sharp image.   Happily, the continuum images had slightly narrower profiles in most cases, so we usually did not have to blur the resolution of the images showing the knots themselves.

Once all the images were precisely registered on a common coordinate system, proper motions from one epoch to another can be seen, either from blinking images from different epochs, or from taking the difference between the two.  In Fig.~1 we show an \oiii\ image of the entire G292 remnant, and in Fig.~2 we show enlarged versions of the difference images between different epochs for the locations indexed on Fig.~1.    From such visual inspection of the images, we selected 67  individual knots or filaments for analysis.  In order to be selected, filaments had to be:   (1) sufficiently isolated that a rectangular box could be placed around the filament, enclosing virtually all of the emission and some surrounding background sky (this was necessary for our correlation analysis, \S3.1); and (2) have no significant trace of residual stars on the filament itself in the continuum-subtracted images.  (Residual stars in the surrounding background sky could be edited out manually, but we were afraid that attempting this on the filaments themselves could bias the data.)  Some filaments were outside the field of view in images from epochs 1986 and/or 1999, or blueshifted out of the pass band in the 1999 and 2000 images, but all appear in images from at least three epochs:  2002, 2006, and 2008.  We found no filaments that either appeared or disappeared over the 21-year span of our observations, unlike the situation with the much younger Cas A \citep{kamper76}.

\section{Proper-Motion Measurements}
All the filaments used in our analysis were selected explicitly because of their strong \oiii \lam 5007
emission and absence of \ha, facts that strongly suggest these are fragments from the progenitor 
core  that have interacted little if any with the interstellar medium since being launched by
the explosion.  It is thus reasonable to expect that the fragments we now see as optical filaments have been coasting with near-constant velocities  since the explosion, a model which we investigate in detail in \S4.  For the moment, however, we describe in this section how we obtain the best values for the motions of the individual filaments.  The only assumption is that the motion of  each filament  has been at essentially constant velocity over the 21-year history of our observations, a period that is less than one percent of the age of the SNR\@.  

The most straightforward approach to the problem might be to measure the position of each filament directly as a function of time using each of the epochs where we have observations.   This is essentially the approach taken by \citet{thorstensen01} for Cas A, who then did a linear fit to measure the proper motions (see \S3.3 for more on this approach).  For G292 this approach has two difficulties: most importantly, virtually all the knots and filaments have an irregular morphology, making it difficult to consistently define a precise center.  This problem is compounded by the second one; namely, that individual filaments may change somewhat over time.  (For example fresh material may encounter the reverse shock and light up, while other material fades.)  \citet{thorstensen01} encountered both these problems in Cas A, but the impact there was less important since the total baseline was 48 years, almost 20\% of the age of the remnant, compared with less than 1\%  in the case of G292.   

In our case the motions are quite subtle; no filament shows a total displacement $>2\arcsec$\  over the course of our observations, and typical epoch-to-epoch displacements are much less---certainly much less than the precision to which we could define the absolute position of a filament at any epoch. (The cases shown in Fig.~2 are among the most extreme.)   But it is not necessary to precisely measure a filament's position on an absolute frame; precisely measuring its {\it shift} in position from  one epoch to another, combined with an approximate measurement of its absolute position at some reference epoch, is sufficient for our purposes.   For the relative position measurements we used a two-dimensional correlation technique described in the next section.

\subsection{Individual Measurements: Two-Dimensional Correlations}
To measure the displacement of an individual filament $\alpha$\ from epoch {\it i} to epoch 
{\it j}, 
we first clipped out identical small rectangular sections from the aligned, continuum-subtracted images at the two epochs.  Typical sections measured 30 to 100 pixels (6\arcsec\ to 20\arcsec) in both {\it x} (E-W) and {\it y} (N-S) directions (exactly the small rectangles shown in Fig.~1), and were chosen to isolate an individual filament.  We then used the IRAF task {\it xregister} to calculate the displacement, ($\Delta x_{\alpha, i, j}, \Delta y_{\alpha, i, j})$\ from epoch {\it i} to {\it j} that gave the best match for the filamentary emission at the two epochs.   The {\it xregister} task is a two-dimensional implementation of the cross-correlation technique described by \citet{tonry79}.  Since the shifts are all $< 2\arcsec$\ (10 pixels), we used a discrete search over a small region, followed by a centroid fit to determine the best fractional-pixel shift in both dimensions.  The cross-correlation technique does not give a reliable uncertainty estimate for an individual measurement;  we use the scatter among measurements derived from multiple epoch pairs to obtain an overall uncertainty for each filament, as described in the next section.

\subsection{Multiple-epoch Measurements: Averages, Weighting, and Uncertainty}
Our next step was to combine all the measurements over multiple baselines to give a mean value for the proper motion  $(\mu _{x_\alpha  } , \mu _{y_\alpha  }) $\ for each of the 67  filaments, along with an uncertainty estimate.  We did this independent of any model assumptions about the long-term kinematics of the filaments; \ie, we did not assume a free-expansion (or any other) model.
In this process we analyzed each transverse dimension, {\it x} and {\it y}, independently;  in what follows we will describe the analysis in the {\it x} direction; that for the {\it y} is, of course, identical.

For each filament \al, we calculate the mean proper motion  $\mu _{x_\alpha  }$\ by taking an average 
of the measurements of $\Delta x_{\alpha, i, j}$, weighted by the square of the 
baseline $\Delta t_{i,j}  \equiv t_j  - t_i $:
\begin{equation}
\mu _{x_\alpha  }  \equiv \overline \mu  _{x_\alpha  }  \equiv \frac{{\sum\limits_{i,j} {\frac{{\Delta x_{\alpha ,i,j} }}{{\Delta t_{i,j} }}\Delta t_{i,j} ^2 } }}{{\sum\limits_{i,j} {\Delta t_{i,j} ^2 } }} = \frac{{\sum\limits_{i,j} {\Delta x_{\alpha ,i,j} \Delta t_{i,j} } }}{{\sum\limits_{i,j} {\Delta t_{i,j} ^2 } }}
\end{equation}
where the shorthand notation $\sum\limits_{i,j}$\ 
is used to indicate a sum taken over all the unique
baselines, $\sum\limits_j {\sum\limits_{i < j}} $\@.  
(To see that $\Delta t^2$\ is the correct weighting, consider that in combining data values with different known signal-to-noise, one weights as $1/\sigma^2$, or $(S/N)^2$\@.  For our correlation measurements, the noise is approximately the same for any epoch pair, but the signal, i.e., the displacement, increases linearly with the baseline.   Hence $S/N \propto \Delta t_{i,j}$, and $\Delta t_{i,j}^2$\ is the best weighting factor.)
For filaments that appear in observations from 
all seven epochs, there are $7\times 6/2 = 21$\ distinct baselines to be considered.  For those 
filaments that are present in images from fewer epochs (because they lie outside the small fields
observed in 1986 and/or 1991, or because they were blueshifted outside the filter bandpass in 1999 and 2000), there are correspondingly fewer baselines.  We shall denote 
the total number of baselines included in the average for each filament as {\it N}.  

The advantage of using this multiple-baseline approach, compared to a simpler approach we describe briefly in \S3.3,  is that we  have a distribution comprising  far more measurements whose variance we can calculate in order to estimate the uncertainties.  
The (unbiased) variance $S_{x_\alpha } ^2$\ in the sample of measurements $\mu  _{x_{\alpha ,i,j}}$\  with weights $\Delta t_{i,j}^2$ is 
\begin{equation}
S_{x_\alpha  } ^2  = \frac{{\sum\limits_{i,j} {\Delta t_{i,j} ^2 } }}{{\left( {\sum\limits_{i,j} {\Delta t_{i,j} ^2 } } \right)^2  - \sum\limits_{i,j} {\Delta t_{i,j} ^4 } }}\sum\limits_{i,j} {\left( {\mu _{\alpha ,i,j}  - \overline \mu  } \right)^2 \Delta t_{i,j} ^2 } 
\end{equation}
and the uncertainty $\sigma (\mu _{x_\alpha  } )$ in the mean measurement $\mu _{x_\alpha  }$\ is best estimated by 
\begin{equation}
\sigma (\mu _{x_\alpha  } )^2  = \frac{{\sum\limits_{i,j} {\Delta t_{i,j} ^4 } }}{{\left( {\sum\limits_{i,j} {\Delta t_{i,j} ^2 } } \right)^2 }}S_{x_\alpha  } ^2 
\end{equation}
\citep[e.g.,][]{bevington02}.

In Fig.~3 we show a plot of individual proper-motion measurements, $\mu_{x_{\alpha,i,j}}$\ and $\mu_{y_{\alpha,i,j}}$, as a function of time difference $\Delta t_{i,j}$\ for all baselines, for three prominent filaments (the same ones highlighted in Fig.~2).   Table 2 gives the measured values of positions and mean motions  $\mu_{x_{\alpha}},\  \mu_{y_{\alpha}}$\ (with uncertainties) for all 67  filaments.  We measured the positions in the 2002 image (the one with the best seeing) and extrapolated back to epoch 2000.0.   We present the same data graphically, as vectors representing the proper motions extrapolated 1000 years into the future at the present rate, superimposed on an image of G292 in Fig.~4\@.  This figure clearly shows, at least qualitatively and as projected onto the plane of the sky, that the filaments are moving outward from a point near the center of G292, and that the ones farthest from the center are moving the fastest.  We explore such a free-expansion model quantitatively in \S4.

We note that the measured proper motions are small, only $20\ {\rm to}\ 133\masy$, with only 4 of the 67 (all in the extreme south) having motions $>100\masy$\@.   This is far smaller than for the closer and much younger oxygen-rich SNR Cas A, where \citet{thorstensen01} measured filamentary proper motions of $\sim160\ {\rm to}\ 800\masy$\@.

\subsection{Alternative Linear-Fit Approach}

An alternative to the multiple-epoch averages just described would be simply to plot the (absolute or relative) position of a filament as a function of time, and determine its motion from the slope of a linear fit.  This is essentially the approach used  for Cas A by \citet{thorstensen01}, who  measured absolute filament positions on a standardized reference grid.  We have tried the same technique using the relative positions, measured as described in \S3.1.     In Fig. 5 we show the results for the same three prominent filaments  as those highlighted in Figs.~2 and 3.   

This linear-fit approach provides a perfectly reasonable measure for the proper motions, but it does not extract the maximum amount of information from the data at hand.  For a filament with observations at 
all seven epochs, the linear-fit method uses only six epoch pairs, compared with 21 for the multiple-epoch averages described in \S3.2\@.  Not surprisingly, the uncertainties with the linear-fit method are generally larger than for the weighted multi-epoch averages (\S3.2).
Nevertheless, we went on to fit both versions of the data using the free-expansion model, and obtained  similar results (\S4).

\section{Results: Global Kinematics and Free-Expansion Model}

The simplest model for the overall kinematics of the filaments in G292 is that the filaments we see today have been coasting with constant velocity since being launched by the explosion, and only recently rendered visible through interaction with a shock.  Such a model has been successfully applied to the other two oxygen-rich SNRs in the Galaxy, Cas A \citep{thorstensen01} and Puppis A \citep{winkler88}.  Here we will use it for G292.   

For undecelerated motion, 
a filament now located at position ${\bf{r}} = (x, y)$\ is 
expected to have proper motion $(\mu _{x_\alpha  } , \mu _{y_\alpha  }) $\  given by 
\begin{equation}
\overrightarrow \mu   \equiv \frac{{d{\bf{r}}}}{{dt}} = \frac{{\left( {{\bf{r}} - {\bf{r}}_0 } \right)}}{\tau }, {\rm or}\ 
\  \mu _{x_\alpha  }  = \frac{{\left( {x_\alpha   - x_0 } \right)}}{\tau } ,\  \ \mu _{y_\alpha  }  = \frac{{\left( {y_\alpha   - y_0 } \right)}}{\tau }\; ,
\end{equation}
where ${\bf{r}}_0 \equiv (x_0, y_0)$\  gives the coordinates of the center of expansion (presumably the location 
of the progenitor at the time it exploded), and $\tau$\ is the age.  Of course $(x_0, y_0)$\  and $\tau$\ 
are parameters of the model.    

Considering the ensemble of all 67 filaments, we have plotted components of the proper motion 
$\left( {\mu _x ,\mu _y } \right)$\ as a function of $(x, y)$, with the results shown in Fig.~6.  The error bars represent the uncertainties as calculated in \S3.2 and given in Table 2\@.  It is obvious that these data are generally consistent with constant-velocity expansion, so we have performed linear fits to obtain values for $x_0$\ (the $x$-intercept) and $\tau$\ (the inverse slope).  
The results of this analysis are:
\begin{equation}
{\rm R.A.}\:(2000.0) = 11^h 24^m 34.4^s,\  {\rm Dec.}\: (2000.0) = -59\degr 15\arcmin 51\arcsec,\  \tau = 2990 \; {\rm yr}.
\end{equation}


Estimating the uncertainties to attach to these values requires more than a formal error-propagation analysis.   Evaluation of  $\chi^2$ for the linear fit (Eq. 4) to the data of Table 2 (and Fig.~6), we find an unacceptably large value:  5.3 per degree of freedom.   (There are $67 \times 2\ {\rm coordinates} - 3\ {\rm parameters}=131$\ degrees of freedom.)  This should not come as a surprise, and does not mean we should reject the model entirely.  There is good reason to believe that free expansion, while giving a good overall description for the data, is not a perfect model.  Individual knots of ejecta become excited, and therefore visible, as they encounter the reverse shock, a process that also leads to the deceleration of these knots.   Furthermore, some new material may become excited while that which has been shocked earlier fades, a process that will also mimic deceleration.   And finally,  although ejecta knots are themselves  moving radially,  the reverse shock may have developed a non-spherical geometry in some directions, so knot-shock encounters can result in apparent changes in direction as well as decelerations for the excited filaments.   All of these effects are seen in G292's ``younger cousin" Cas A, where the velocities are up to 5 times faster, and the age 10 times younger \citep{fesen01a,thorstensen01,fesen06b}.

While it remains clear that the constant-expansion model is a good overall description for the O-rich filaments in G292, the large $\chi^2$ values mean that we cannot reliably use incremental contours in $\chi^2$ space to determine confidence limits on the model parameters.  As an alternative we turn to the bootstrap technique described in \citet{press07}\@.  Briefly, in this technique one starts with the original sample of {\it M} points (in our case, the 67 filaments), and randomly selects exactly {\it M} values, {\it with replacement}.  Thus some points may be selected multiple times; others not at all.   We fit the new data sample with the same 3-parameter model (Eq. 4) as the real data  to determine $(x_0, y_0)$\  and $\tau$.  We then repeat this process for 100,000 trials, and consider the resulting ensemble of model parameters.  The mean values from the 100,000 trials are in extremely close agreement with those from the unique best fit to the actual data, but now we can examine the frequency to give a robust estimate for the confidence limits.  

The resulting allowed region for the expansion center is shown in Fig.~7, an enlargement of the central region of G292.  The 90\%-confidence contours measure approximately $\Delta {\rm (R.A.)} = ^{+4.2\arcsec}_{-3.2\arcsec}, \Delta {\rm (Dec.)} = \pm 4.8\arcsec$\@.   The uncertainty in age is $\pm 60\; {\rm yr}\ (1\sigma),\ {\rm or}\ ^{+90}_{-100}\; {\rm yr}$\ (90\%-confidence).


We have investigated the differences between our measured values for the proper motions and the best-fit free-expansion model to check for a possible systematic pattern.  In Fig.~8 we show the measured values as red vectors (identical to those in Fig.~4), and the corresponding proper motions from the model in blue.  The uncertainties are indicated as ellipses at the ends of the measured-value vectors.  While there are a number of cases where the deviation between measured and model is large compared with the uncertainty (thus producing the high $\chi^2$\ value), we can discern no pattern to these.  In several of these cases, the error ellipse is highly elongated, i.e., the formal uncertainty is much higher in one coordinate than the other.  In at least some of them, the small uncertainty in one direction may have resulted from a fortuitously low dispersion among a small number of measurements.

We have also carried out fits using the average proper-motion values calculated using the simpler linear-fit method of \S3.3, and found  very similar  results:  an expansion center within 2\arcsec\ of that given in Eq.~5, with somewhat larger but similarly shaped error contours.   The age resulting from these fits was $2900\pm 80 \; {\rm yr}\; (1\sigma)$, overlapping with the value from Eq.~5 at the $1\sigma$\ level.  Given the great similarity in the two sets of results, we shall use only those from our preferred multi-epoch averages (Eq.~5) in the remainder of this paper.


\section{Discussion}
G292 shows a somewhat irregular morphology at all wavelengths.  
The largest feature, and the one which comes closest to circular symmetry, is the outer ``plateau" region, as shown in the 20-cm radio image taken from the Australia Telescope Compact Array by \citet{gaensler03}.  This shows a sharp outer shell with radius $\sim 4\arcmin$, centered at 
${\rm R.A.}\:(2000.0) = 11^h 24^m 34.8^s,\  {\rm Dec.}\: (2000.0) = -59\degr 15\arcmin 52.9\arcsec$, with an uncertainty of $\pm 5\arcsec$\ in both directions.   The same image shows that the brightest radio emission is concentrated within a much brighter inner core of radius $\sim 2\arcmin$, with a center displaced  $\sim 30\arcsec$\ east and south from that of the outer shell.  \citet{gaensler03} identify the core as a pulsar-wind nebula associated with the PSR J1124--5916 (see below), and the plateau boundary as the outer shell where the supernova blast wave is interacting with its environment.

Our measured expansion center (Eq.~5) is remarkably close to the center of this outer radio shell: only 3\arcsec\ away, as shown in Fig.~7\@.  While such close agreement (within the uncertainties of both center determinations) may be fortuitous, this common center makes it appear likely that the outer blast wave from the explosion that has shaped the radio shell has approximate spherical symmetry.   Certainly it is more symmetrical than the distribution of inner ejecta that we are now seeing as \oiii-emitting filaments (see subsequent discussion), or than the bright core of radio emission, or than the X-ray-emitting plasma, which comprises material with varying degrees of enrichment from the supernova ejecta \citep{park04, park07}.   


The young pulsar PSR J1124--5916, located near the center of G292, is almost certainly the compact remnant from the same supernova \citep{camilo02}.  This has already been underscored by the presence of a prominent pulsar-wind nebula in X-rays, discovered by \citet{hughes01} prior to the discovery of the relatively faint pulsar.  \citet{camilo02} have measured the spin-down age of the pulsar at 2900 yr, quite close to the 2990-year expansion age for the G292 remnant and further cementing their association.  

The pulsar is now located at 
${\rm R.A.}\:(2000.0) = 11^h 24^m 39.1^s,\  {\rm Dec.}\: (2000.0) = -59\degr 16\arcmin 20\arcsec$, 46\arcsec\ southeast of the expansion center (Eq.~5).   If we assume that the pulsar and the oxygen-rich filaments were launched at the same time, about 3000 years ago, we would expect the pulsar to be moving southeastward at $16 \masy$, or a transverse space velocity of $440\: d_6 \kms$\@.  While fast to be sure, this is well within the normal range for pulsar birth velocities \citep{caraveo93, frail94}, and far short of the transverse velocity of the unpulsed neutron star RX J0822--4300 inside the oxygen-rich SNR Puppis A, directly measured (using X-ray data from {\it Chandra}) to be at least $1000\kms$\ \citep{hui06a}, and probably closer to $1600 \kms$\ \citep{winkler07}\@.   A similar direct proper-motion measurement for the pulsar in G292 will present an extreme challenge.   Nevertheless, the displacement of the pulsar from the expansion center of the remnant is further evidence that core-collapse supernovae have significant asymmetries.

\citet{ghavamian05} used imaging Fabry-Perot spectroscopy to explore the kinematics of G292 through measurement of radial velocities for the \oiii-emitting filaments in the central region of G292---a field including all but the filaments most distant from the center (and thus with the highest transverse velocity).     They found at least some filaments over the entire velocity range they investigated: $-1440$\ to $+1700 \kms$.  They plotted the radial velocity of the filaments as a function of their projected distance from the geometric center of the remnant, measured by them from the \citet{gaensler03} radio images as a point about 22\arcsec\ north of the center we determined.  These data show that the ejecta filaments are irregularly distributed  around a thick ellipsoidal shell in the position-radial velocity plane, suggesting a shell expanding with a speed of $\sim 1700\kms$.  A simple scaling argument, assuming that the shell is expanding with transverse expansion velocity equal to their observed radial velocity extremes, led them to an age estimate of (3000 - 3400)$\;d_6$\ yr.  

Our own measurement of the age, based as it is on actual measurement of the transverse motions of the filaments rather than estimating them from a model, is both more reliable and more precise, and is also independent of the distance to G292.  The fact that the \citet{ghavamian05} age measurement is roughly consistent with ours indicates that the distance of 6 kpc \citep{gaensler03} is probably about right.  

Taken together, the proper-motion and radial-velocity measurements indicate a definite asymmetry in the distribution of ejecta-dominated filaments that are now visible in G292.  The proper-motion measurements, and also the spatial distribution of oxygen-rich filaments, show that the ejecta along an axis oriented roughly N-S are moving significantly faster than those along the E-W axis.   In the E-W direction, we find $-63 < \mu_x < 52\masy$, or $-1800\: d_6 < v_x < 1490\: d_6\kms$, compared with  $-125 < \mu_y < 82 \masy$, or $-3570\: d_6 < v_y < 2340\: d_6 \kms$\ in the N-S direction.  Furthermore, \citet{ghavamian05} measured radial velocities $- 1440\kms \lesssim v_{rad} \lesssim 1700\kms$, virtually identical with the E-W range we have measured, so it is clear that the fastest moving visible filaments are those moving north and south.\footnote{\citet{ghavamian05} observed some filaments throughout the radial velocity range they examined, so one might wonder if there could be even faster filaments.   However, in comparing their Fabry-Perot images  with our direct ones, which have a bandwidth extending to $\pm 1800\kms$\ (FWHM, and beyond at decreasing efficiency), we find no filaments in the central region of G292 that are not also present in the \citet{ghavamian05} images.  It is highly unlikely that filaments with higher radial velocities are lurking unseen in G292.}     The most straightforward interpretation of these results is that the current distribution reflects an asymmetry in the supernova itself, and that core material was  ejected fastest along an axis oriented approximately N-S in the plane of the sky.   However, the alternative of circumstellar material, distributed roughly in a plane perpendicular to the N-S axis and shaping the flow of ejecta, cannot be excluded.  
An ejecta distribution similar to that in G292 is found in more extreme form in Cas A, 
where the fastest knots are located in a prominent jet to the NE, with a less obvious jet to the SW \citep{fesen88, fesen96, fesen01}.   Asymmetric, possibly bipolar explosions seem likely to have produced both of these oxygen-rich remnants, and perhaps this is a general feature of core-collapse supernovae.   

With proper-motion measurements for the oxygen-rich filaments in hand, it will be possible to combine these with radial velocity measurements to determine a fuller 3-dimensional model for the distribution of the visible ejecta in G292.  We have recently obtained  spectra of dozens of filaments that can be used in conjunction with Fabry-Perot images to pursue this project, but this analysis is left until a subsequent paper.

\acknowledgments

We gratefully acknowledge the outstanding support,  
typical of the mountain staff at CTIO, during the observations from numerous observing 
runs over two decades that yielded the 
data for this analysis.    We have benefitted from discussions with John Thorstensen on the subtleties of measuring the kinematics of young SNRs, a problem we have all wrestled with.
This work has been supported financially by  
the NSF, through grant 
AST-0307613 to P.F.W., and from NASA, through grant NAG 5-8020 to P.F.W. and 
{\it Chandra} grants GO0-1120X and GO1-2058A to K.S.L.  

\bibliographystyle{apj}

\clearpage

\begin{deluxetable}{lccccccrrl}
\tabletypesize{\scriptsize}
\tablewidth{0pt}
\tablecaption{Journal of CTIO Observations}

\tablehead{
\colhead {} &\colhead {} & \colhead {} & \colhead{Scale} &\colhead{} &
\multicolumn{3}{c}{Filter} & 
\colhead{} & 
\colhead {}\\ 

\cline{6-8}  

\colhead{Date (U.T.)} & 
\colhead {Telescope} &
\colhead {CCD} &
\colhead {(\arcsec \thinspace pixel$^{-1}$)} & 
\colhead{ Field (\arcmin)} &
\colhead{Line} &
\colhead{$\lambda_c\;$(\AA)} &
\colhead{$\Delta \lambda$\tablenotemark{a}(\AA)} &
\colhead {Exposure (s)} &
\colhead {Observers} 
}

\startdata

1986 Nov 29 &4.0 m & RCA & 0.589 & 5.0$\times$2.9 & [O III] & 5020 &54\phn\phn & $2\times1000$\phn\phn & PFW, R.P. \\
&&&&&&&&&Kirshner,\\
&&&&&&&&&J.P. Hughes\\
&&&&&&&&&\\
1991 Apr 19 & 4.0 m &TEK 1K \#1&0.470&8.0&[O III]&5020&54\phn\phn&$5\times600$\phn\phn&PFW, KSL \\
&&&&&Blue&4770&100\phn\phn&4$\times$600\phn\phn&\\
&&&&&&&&&\\
1999 Jan 20-23 &0.9 m&TEK 2K \#3&0.401&13.7&[O III]&5026&39\phn\phn&$3 \times1200$\phn\phn&PFW, D. Paul\\
&&&&&Green&5135&90\phn\phn&$3\times 600$\phn\phn&\\\
&&&&&&&&&\\
2000 Jan 30-31&0.9 m&TEK2K \#3&0.401&13.7&[O III]&5026&39\phn\phn&$2 \times1200$\phn\phn &PFW, KSL, \\
&&&&&&&&$1 \times 1000$\phn\phn&E. Galle\\
&&&&&Green&5135&90\phn\phn&$2 \times 600\phn\phn$&\\
&&&&&&&&$1 \times 500$\phn\phn&\\
&&&&&&&&&\\
2002 Mar 20-23&0.9 m&TEK 2K \#3&0.401&13.7&[O III]&5006&60\phn\phn&$5 \times 1000$\phn\phn&PFW, KSL, \\
&&&&&Green&5135&90\phn\phn&$5 \times 500$\phn\phn&C. Reith\\
&&&&&&&&&\\
2006 Mar 30, &0.9 m&TEK 2K \#3&0.401&13.7&[O III]&5006&60\phn\phn&$4 \times 1000$\phn\phn &PFW, KSL, \\
Apr 2 &&&&&&&&$3 \times 800$\phn\phn&K. Twelker \\
&&&&&Green&5135&90\phn\phn&$4 \times 500$\phn\phn &\\
&&&&&&&&$3 \times 400$\phn\phn&\\
&&&&&&&&&\\
2008 Mar 4-6&0.9 m&TEK 2K \#3&0.401&13.7&[O III]&5006&60\phn\phn&$5 \times 1000$\phn\phn&PFW, KSL\\
&&&&&Green&5135&90\phn\phn&$5 \times 500$\phn\phn&\\

\enddata

\tablenotetext{a}{Full width at half maximum in the telescope beam.}

\end{deluxetable}

\begin{deluxetable}{rrrrrrrc}
\tabletypesize{\small}
\tablewidth{0pt}
\tablecaption{Proper Motions of Individual Filaments in the SNR G292.0+1.8}

\tablehead{
\colhead {Fil. No.}  & 
\colhead {R.A.\tablenotemark{a}}  &  
\colhead {Dec.\tablenotemark{a}}  &  
\colhead{$\mu_{\alpha}$}  & 
\colhead{$\sigma_{\mu_{\alpha}}$}  &  
\colhead{$\mu_{\delta}$}  & 
\colhead{$\sigma_{\mu_{\delta}}$}  &  
\colhead {No. Epochs}\\ 

\colhead{}  &  
\colhead {$\arcsec$}  & 
\colhead {$\arcsec$}  & 
\colhead{$\rm{mas\:yr^{-1}}$}  &  
\colhead{$\rm{mas\:yr^{-1}}$}  &  
\colhead{$\rm{mas\:yr^{-1}}$}  &  
\colhead{$\rm{mas\:yr^{-1}}$}  &  
\colhead{} 
}

\startdata
			1 & -137.7 & -112.9 & -64.9 & 16.6 & -18.5 & 3.4 & 6 \\ 
			2 & -90.2 & 6.4 & -60.0 & 7.4 & 12.6 & 3.5 & 4 \\ 
			3 & -90.0 & -157.4 & -52.0 & 4.9 & -54.7 & 4.9 & 5 \\ 
			4 & -89.4 & 53.5 & -34.4 & 10.0 & 24.7 & 5.7 & 5 \\ 
			5 & -85.6 & -130.8 & -31.8 & 7.0 & -49.5 & 7.3 & 6 \\ 
			6 & -83.8 & -185.9 & -32.1 & 5.6 & -57.7 & 3.1 & 6 \\ 
			7 & -83.0 & -193.9 & -32.8 & 3.8 & -58.1 & 2.6 & 5 \\ 
			8 & -82.2 & -291.8 & -24.1 & 3.9 & -79.9 & 9.3 & 6 \\ 
			9 & -65.5 & -290.6 & -42.2 & 4.9 & -122.1 & 8.1 & 5 \\ 
			10 & -64.8 & -300.2 & -44.0 & 5.6 & -105.5 & 7.6 & 5 \\[8 pt] 
			11 & -63.1 & -17.9 & -26.9 & 8.0 & -19.2 & 10.1 & 4 \\ 
			12 & -53.5 & -126.4 & -24.0 & 3.2 & -26.5 & 9.0 & 6 \\ 
			13 & -51.3 & -196.3 & -20.9 & 4.2 & -61.2 & 8.4 & 6 \\ 
			14 & -48.9 & 125.3 & -19.7 & 2.6 & 46.7 & 2.9 & 6 \\ 
			15 & -32.8 & 98.4 & -4.2 & 3.9 & 42.8 & 3.7 & 6 \\ 
			16 & -32.2 & 22.8 & -20.0 & 0.5 & 9.5 & 1.6 & 4 \\ 
			17 & -28.6 & -135.6 & -14.5 & 7.6 & -22.5 & 4.3 & 5 \\ 
			18 & -28.4 & -305.4 & -14.9 & 3.9 & -101.2 & 8.8 & 5 \\ 
			19 & -28.0 & 13.8 & -15.6 & 4.3 & 24.8 & 2.4 & 4 \\ 
			20 & -15.3 & -304.8 & -25.0 & 6.6 & -107.1 & 8.3 & 5 \\[8 pt] 
			21 & -11.6 & -120.3 & -18.1 & 6.1 & -19.4 & 7.6 & 4 \\ 
			22 & -7.4 & 91.6 & -13.6 & 3.6 & 35.1 & 3.3 & 4 \\ 
			23 & 6.7 & 52.2 & -8.1 & 0.7 & 19.7 & 3.7 & 5 \\ 
			24 & 7.7 & 113.7 & -17.0 & 2.1 & 23.8 & 6.0 & 6 \\ 
			25 & 7.7 & 175.4 & 2.3 & 1.9 & 62.4 & 2.8 & 4 \\ 
			26 & 9.7 & 72.3 & -8.1 & 2.0 & 35.0 & 3.5 & 6 \\ 
			27 & 9.9 & 121.1 & -7.8 & 3.1 & 40.8 & 4.5 & 4 \\ 
			28 & 15.8 & 109.8 & 2.6 & 1.9 & 39.2 & 2.0 & 4 \\ 
			29 & 19.7 & 86.8 & 3.2 & 3.9 & 32.2 & 5.7 & 6 \\ 
			30 & 35.2 & 62.3 & 12.6 & 5.7 & 32.2 & 2.6 & 5 \\[8 pt] 
			31 & 36.9 & 140.5 & 5.6 & 1.0 & 54.8 & 4.8 & 4 \\ 
			32 & 55.4 & 171.2 & 23.3 & 2.4 & 56.4 & 4.7 & 6 \\ 
			33 & 55.8 & 194.8 & 12.9 & 2.7 & 81.0 & 2.5 & 6 \\ 
			34 & 67.4 & 174.6 & 7.1 & 3.3 & 64.0 & 8.0 & 4 \\ 
			35 & 67.8 & -251.7 & 33.4 & 13.6 & -91.1 & 5.7 & 3 \\ 
			36 & 73.4 & -260.9 & 9.1 & 4.1 & -78.6 & 6.8 & 5 \\ 
			37 & 78.6 & -226.0 & 27.4 & 3.1 & -82.5 & 6.5 & 3 \\ 
			38 & 83.6 & -200.1 & 20.3 & 2.2 & -61.8 & 2.5 & 6 \\ 
			39 & 89.5 & -45.1 & 20.6 & 1.0 & -4.2 & 1.3 & 7 \\ 
			40 & 90.1 & -31.7 & 15.4 & 2.0 & -11.8 & 2.3 & 7 \\[8 pt] 
			41 & 93.8 & -121.6 & 20.5 & 4.2 & -25.2 & 3.9 & 6 \\ 
			42 & 94.5 & -60.8 & 26.4 & 2.3 & -16.9 & 1.1 & 7 \\ 
			43 & 96.3 & -21.2 & 21.1 & 1.5 & 6.1 & 1.2 & 7 \\ 
			44 & 97.6 & 146.3 & 29.0 & 5.9 & 67.0 & 8.2 & 4 \\ 
			45 & 104.7 & -6.3 & 25.4 & 1.7 & 1.9 & 1.5 & 7 \\ 
			46 & 105.7 & -31.3 & 22.5 & 1.9 & -0.8 & 1.6 & 7 \\ 
			47 & 106.3 & -37.0 & 26.5 & 1.3 & -7.0 & 1.4 & 7 \\ 
			48 & 106.5 & -52.4 & 20.1 & 2.5 & -10.2 & 0.8 & 7 \\ 
			49 & 107.5 & -78.6 & 24.1 & 1.8 & -16.5 & 1.9 & 7 \\ 
			50 & 112.0 & -28.0 & 27.4 & 1.1 & -8.6 & 1.9 & 7 \\ 
			51 & 113.7 & -108.1 & 30.9 & 5.6 & -27.6 & 3.0 & 6 \\[8 pt] 
			52 & 118.7 & -14.5 & 40.3 & 2.4 & 10.0 & 1.4 & 7 \\ 
			53 & 122.3 & -97.5 & 34.5 & 2.6 & -25.9 & 2.0 & 7 \\ 
			54 & 123.7 & -33.5 & 32.6 & 3.2 & 2.7 & 1.9 & 7 \\ 
			55 & 125.4 & -7.0 & 33.9 & 1.4 & 0.2 & 1.5 & 7 \\ 
			56 & 127.4 & 6.0 & 36.7 & 1.7 & 11.6 & 2.0 & 7 \\ 
			57 & 131.8 & 12.2 & 32.5 & 1.3 & 19.3 & 2.0 & 7 \\ 
			58 & 132.5 & -18.1 & 40.8 & 1.9 & -0.6 & 2.1 & 7 \\ 
			59 & 135.0 & -50.8 & 35.0 & 2.6 & -10.2 & 1.5 & 7 \\ 
			60 & 135.8 & -11.3 & 37.9 & 1.0 & 3.0 & 1.5 & 7 \\[8 pt] 
			61 & 137.0 & -46.0 & 43.6 & 2.5 & -15.4 & 1.8 & 7 \\ 
			62 & 139.4 & 63.1 & 26.2 & 1.9 & 24.5 & 0.9 & 7 \\ 
			63 & 142.2 & 73.6 & 33.5 & 9.8 & 35.2 & 6.9 & 6 \\ 
			64 & 144.3 & 7.2 & 50.0 & 3.3 & 16.8 & 2.7 & 7 \\ 
			65 & 153.0 & -13.2 & 48.3 & 2.5 & 4.4 & 0.7 & 7 \\ 
			66 & 159.9 & 48.8 & 48.6 & 3.8 & 18.5 & 2.1 & 5 \\ 
			67 & 161.0 & 58.5 & 51.1 & 3.3 & 27.6 & 1.8 & 7

\enddata

\tablenotetext{a}{Offsets in arcsec relative to R.A.(J2000.) = 11 24 31.0, Dec.(J2000.) = -- 59 15 30\ (an arbitrarily chosen point near the remnant canter).  Positive offsets are east and north.}

\end{deluxetable}

\clearpage 

\begin{figure}
\epsscale{0.9}
\plotone{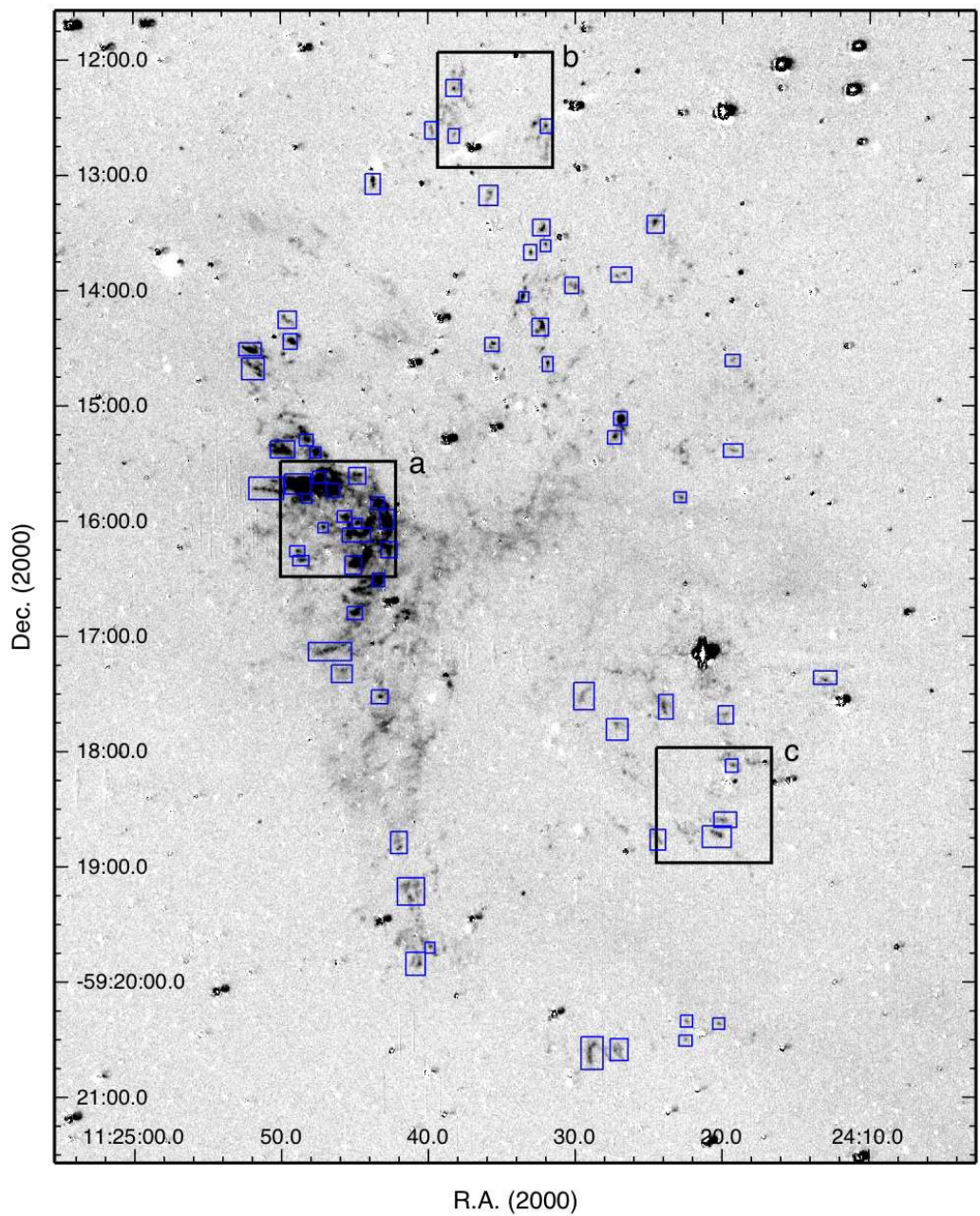}
\caption{Continuum-subtracted image of G292.0+1.8 in the light of \oiii\lam 5007\@.  The small blue boxes indicate the locations of 67 filaments whose proper motions have been measured.  The larger boxes show the regions that appear in the enlarged difference images in Fig.~2\@.  In order to show the fainter filaments clearly,  many of the bright filaments in the eastern ``spur''  region (within and near box {\it a}) are saturated in this display, and may appear to blur together.  In fact, each of the filaments we have used for our measurements (those in the small blue boxes) are distinct.}
\end{figure}

\begin{figure}
\epsscale{1.0}
\plotone{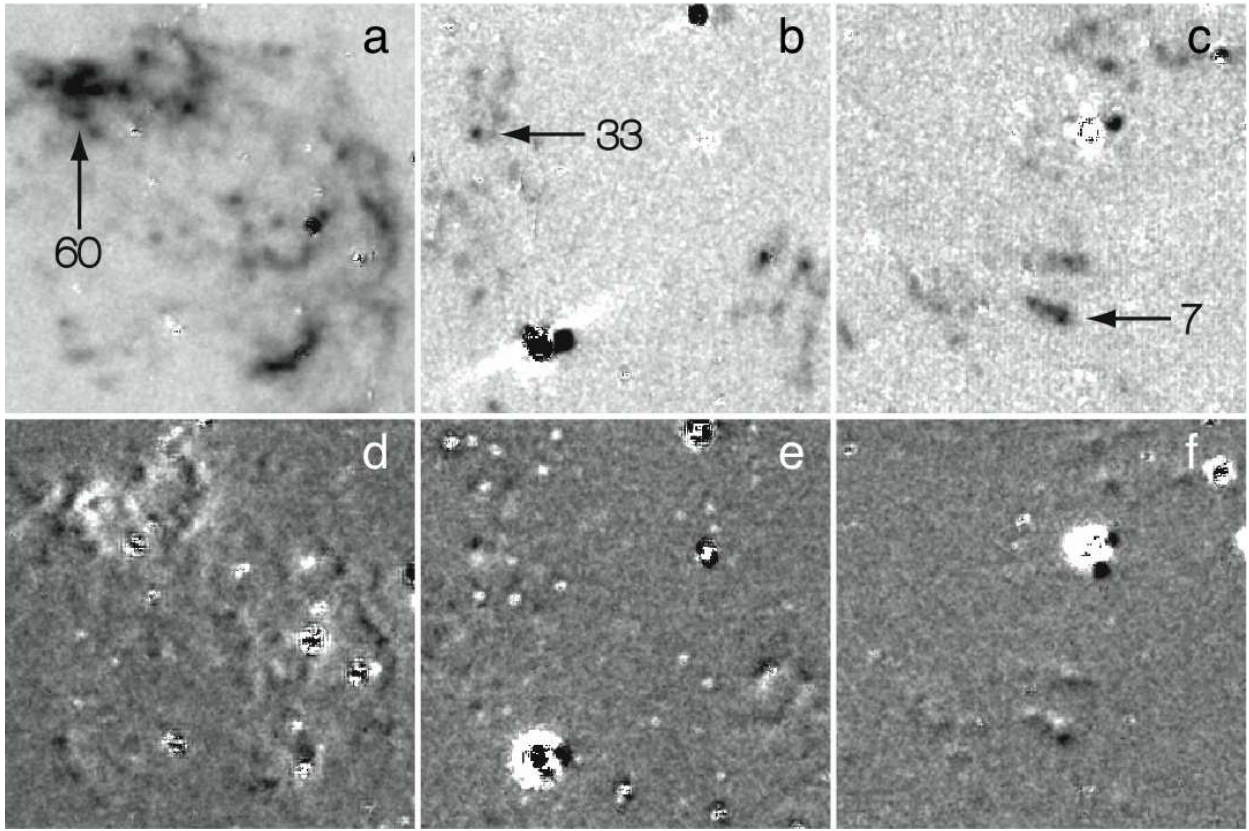}
\caption{The top panels ({\it a - c}) are enlargements of the three 1\arcmin-square sections indicated in Fig. 1, with one particularly prominent filament in each indicated by the arrow.  We use these same filaments to illustrate our analysis  methods in subsequent figures.  The lower panels show the same sections as the difference between images taken at different epochs.  In ({\it d}) the difference is between 1986 and 2008 images, while ({\it e}) and ({\it f}) use the difference between 1991 and 2008 images.  (We have used the longest baseline available showing these filaments.)  In each case emission at the  earlier epoch appears as white, while the later epoch appears as black.  Outward motion, while subtle, is apparent in each case.}
\end{figure}

\begin{figure}
\plottwo{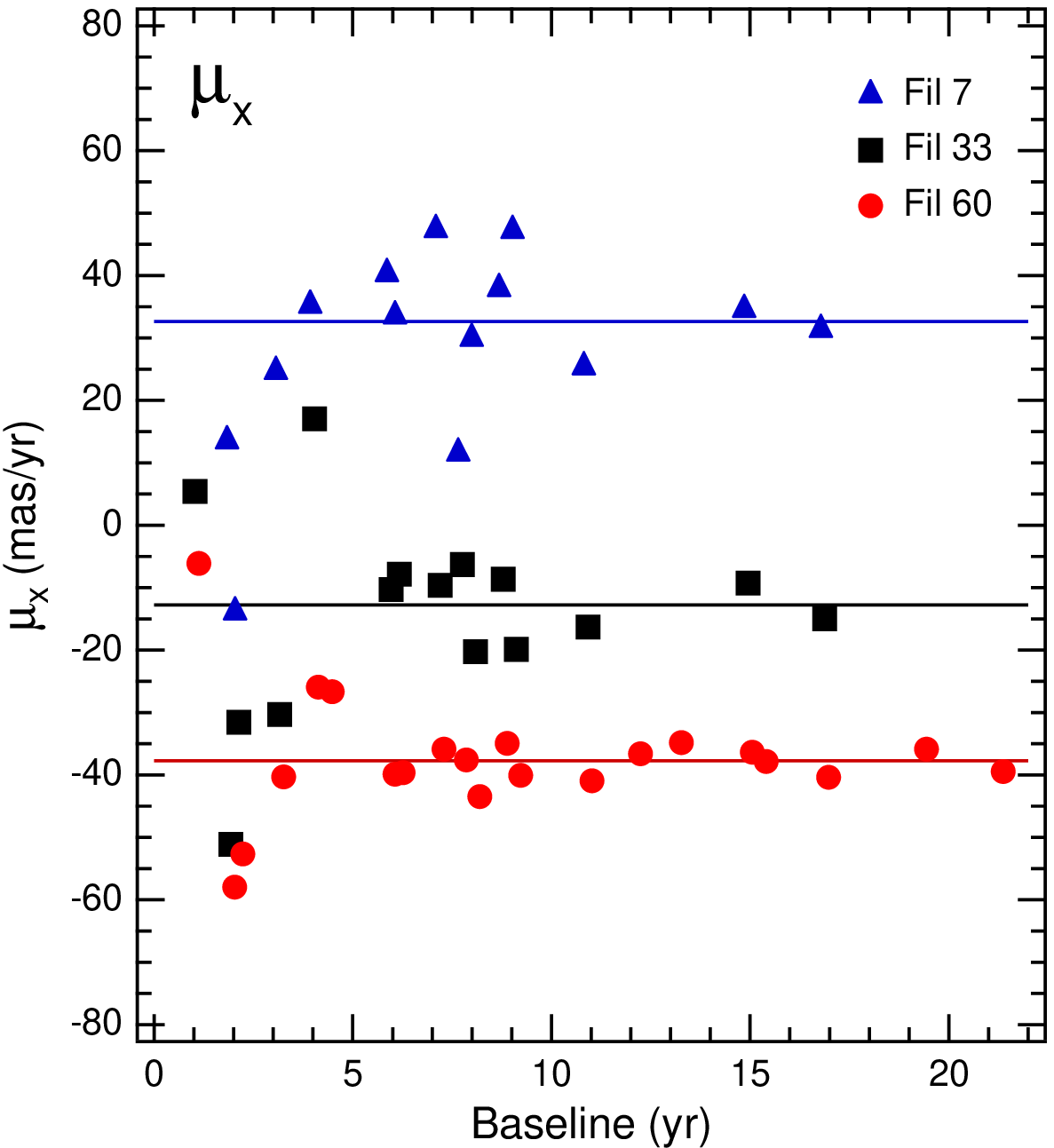}{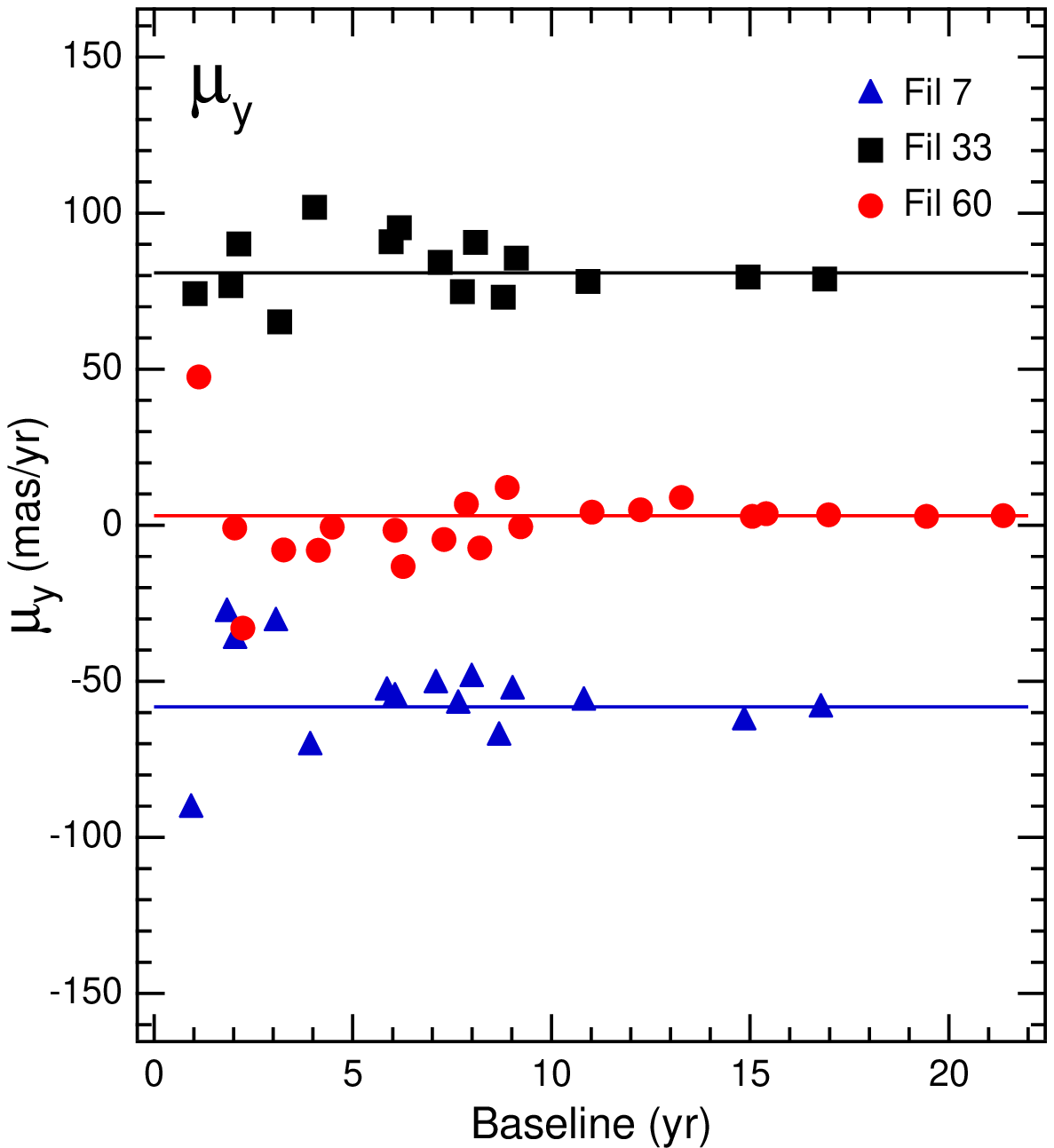}
\caption{Proper motion velocities (in {\it x} and {\it y}) of three prominent filaments.  (These are the same filaments identified in Fig. 2.)   Each epoch pair (21 pairs for filament 60, 15 for filaments 7 and 33) gives an independent measure of the motion.  The dispersion is smallest for the measurements with the longest baselines, and naturally we weight these more heavily in determining the average of all measurements 
(solid lines) which we take as the best overall statistic (\S 3.2).}

\end{figure}

\begin{figure}
\epsscale{0.95}
\plotone{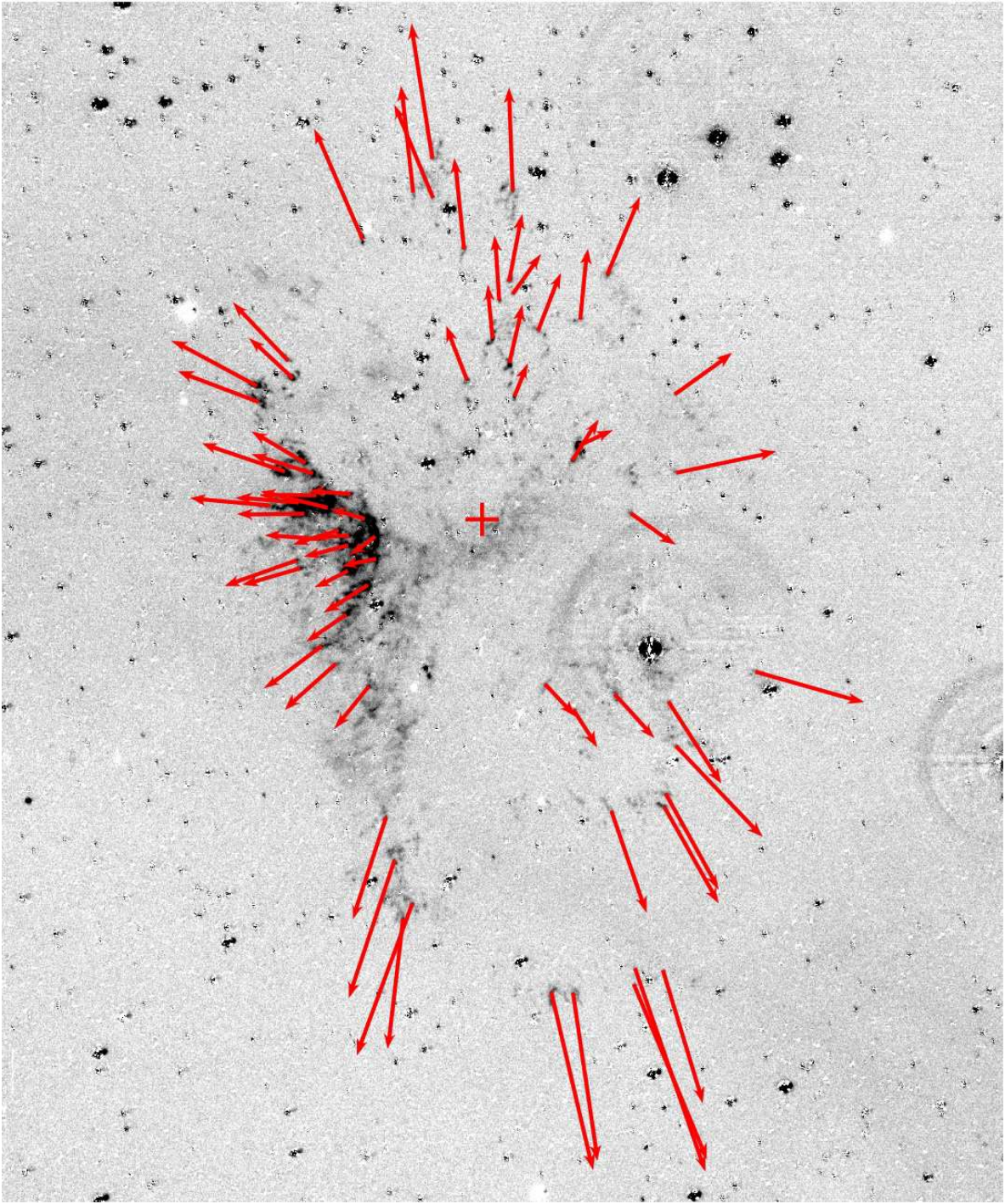}
\caption{Continuum-subtracted image of G292, with vectors representing the measured proper motions of oxygen-rich filaments projected forward 1000 years at the rates we have determined.  The vectors clearly radiate outward from a point near the center of the remnant, indicated by the cross  (see \S 4; the central region is enlarged in Fig.~7).}
\end{figure}

\clearpage

\begin{figure}
\epsscale{0.90}
\plottwo{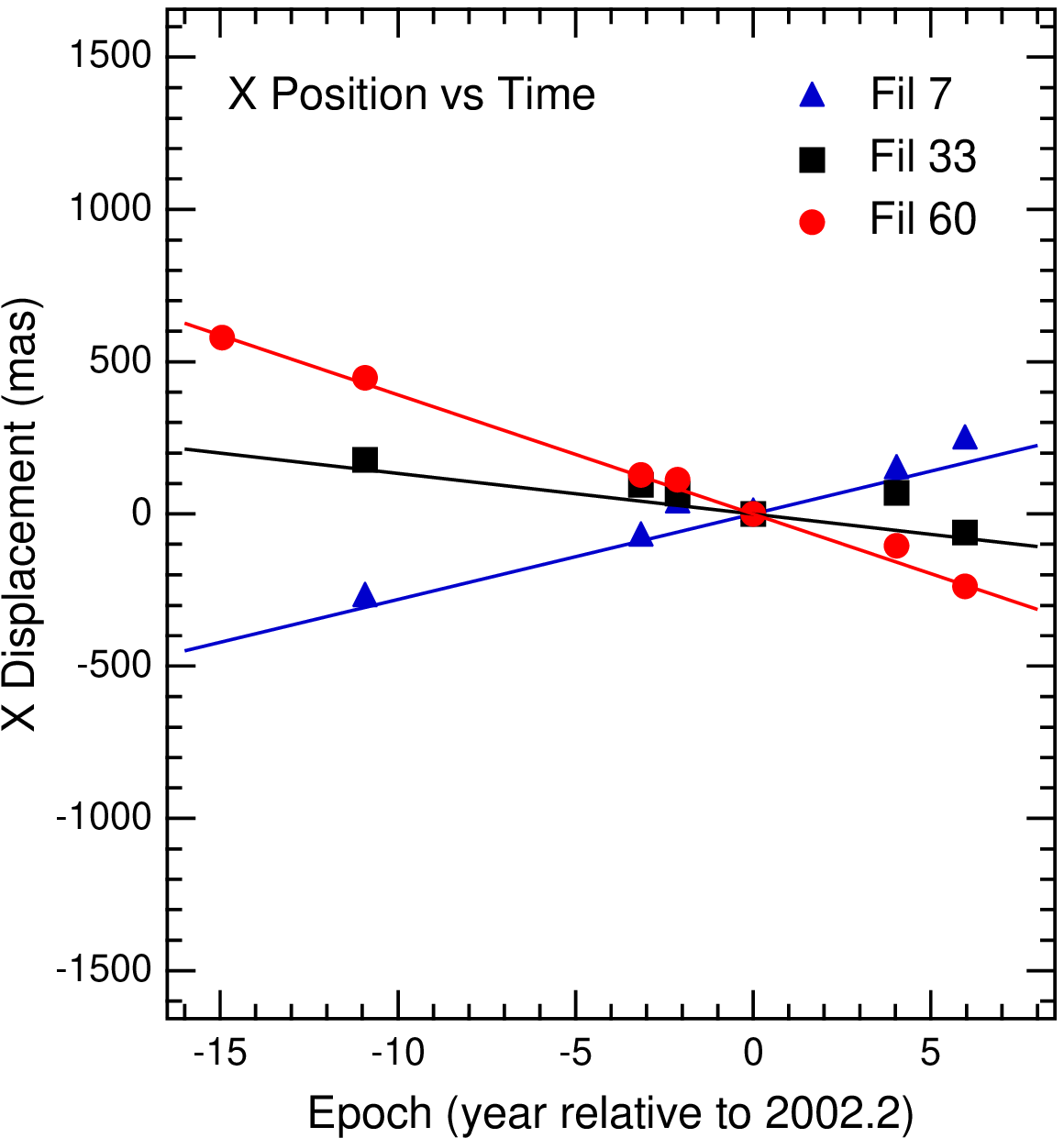}{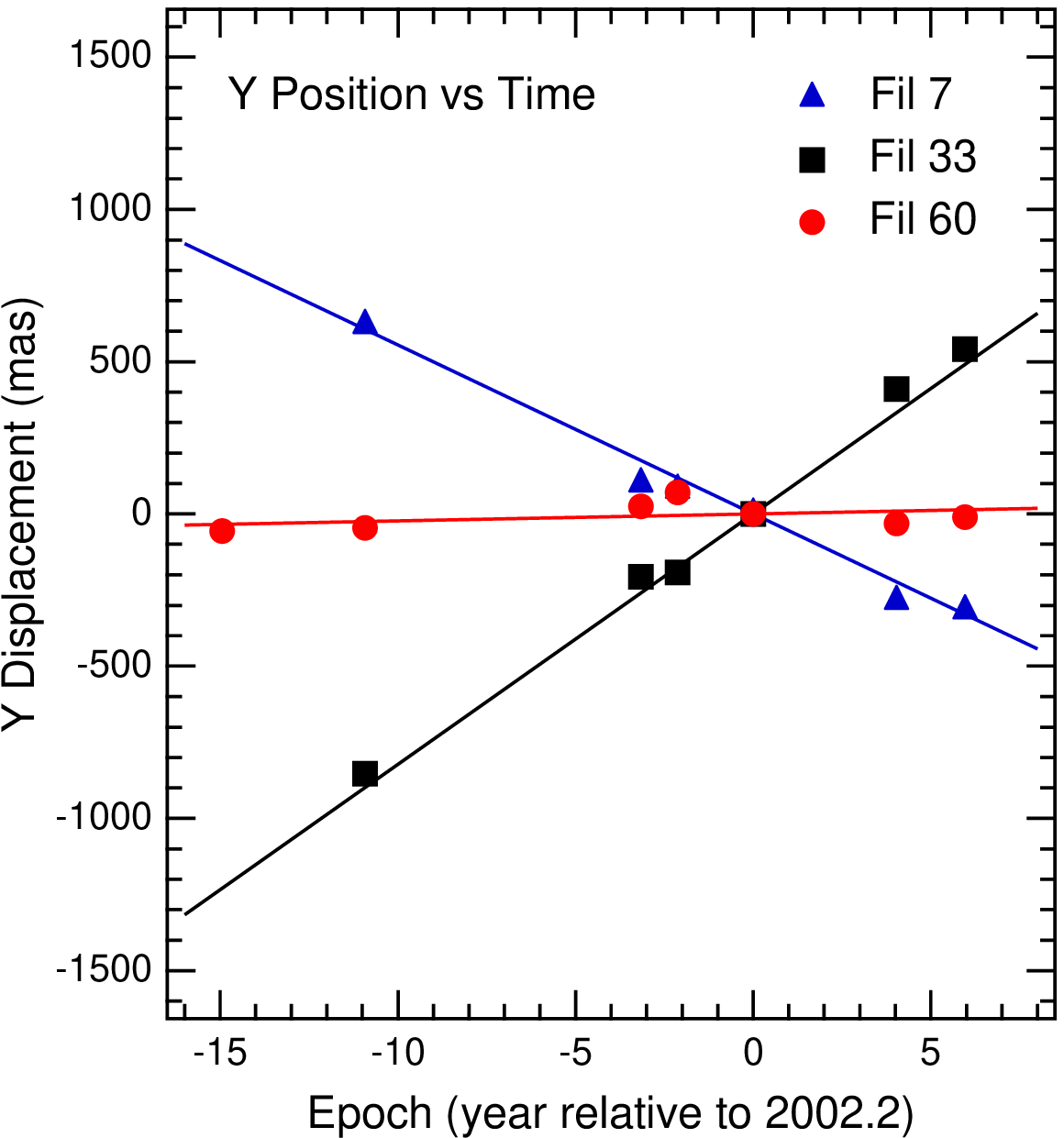}
\caption{Displacements (in {\it x} and {\it y}) of three prominent filaments (the same filaments as in Figs. 2 and 3) relative to their position at our 2002 epoch.  The slope of the best linear fit gives a measure of the proper motion, as described in \S 3.3}
\end{figure}

\begin{figure}
\epsscale{0.90}
\plottwo{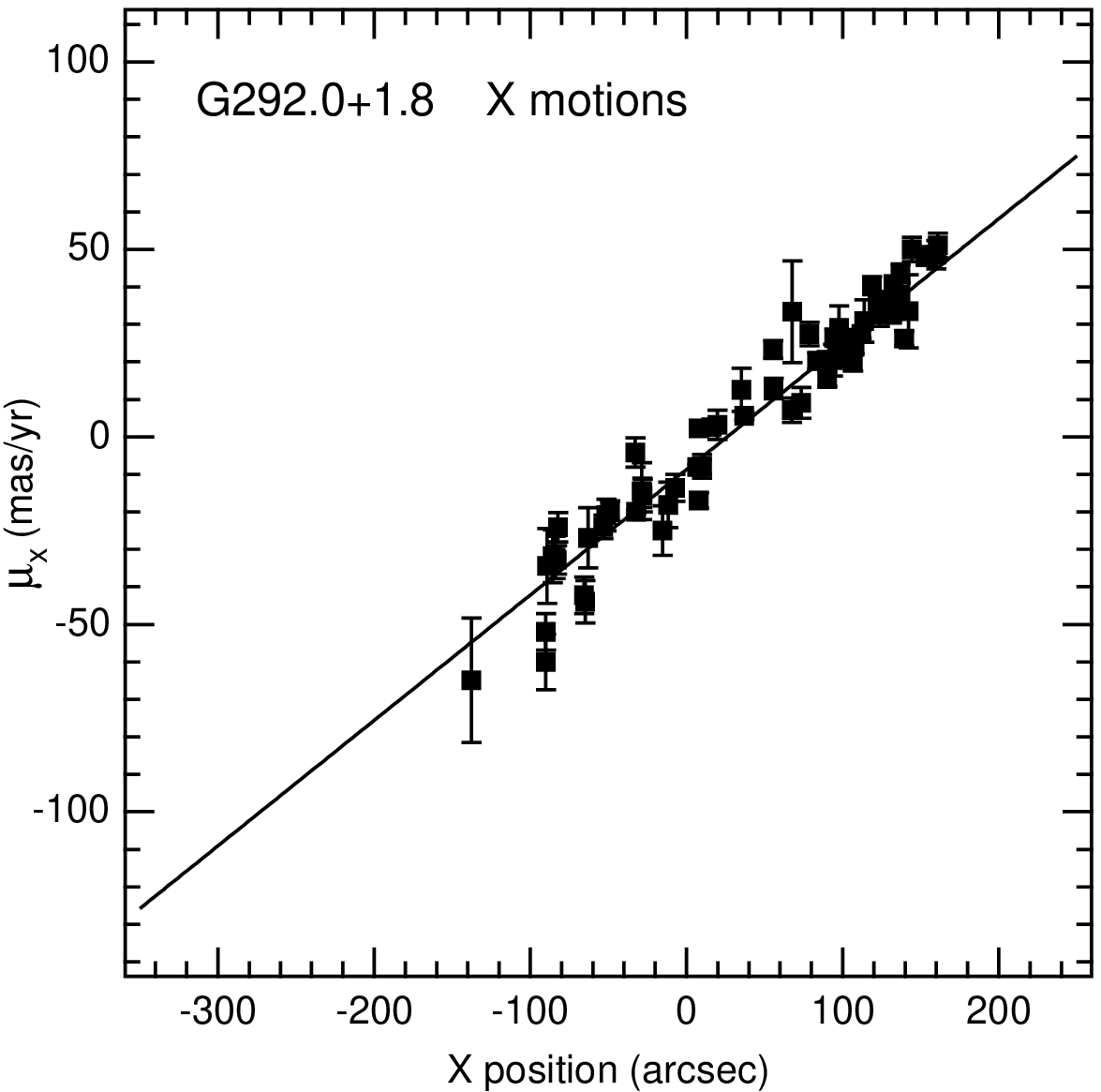}{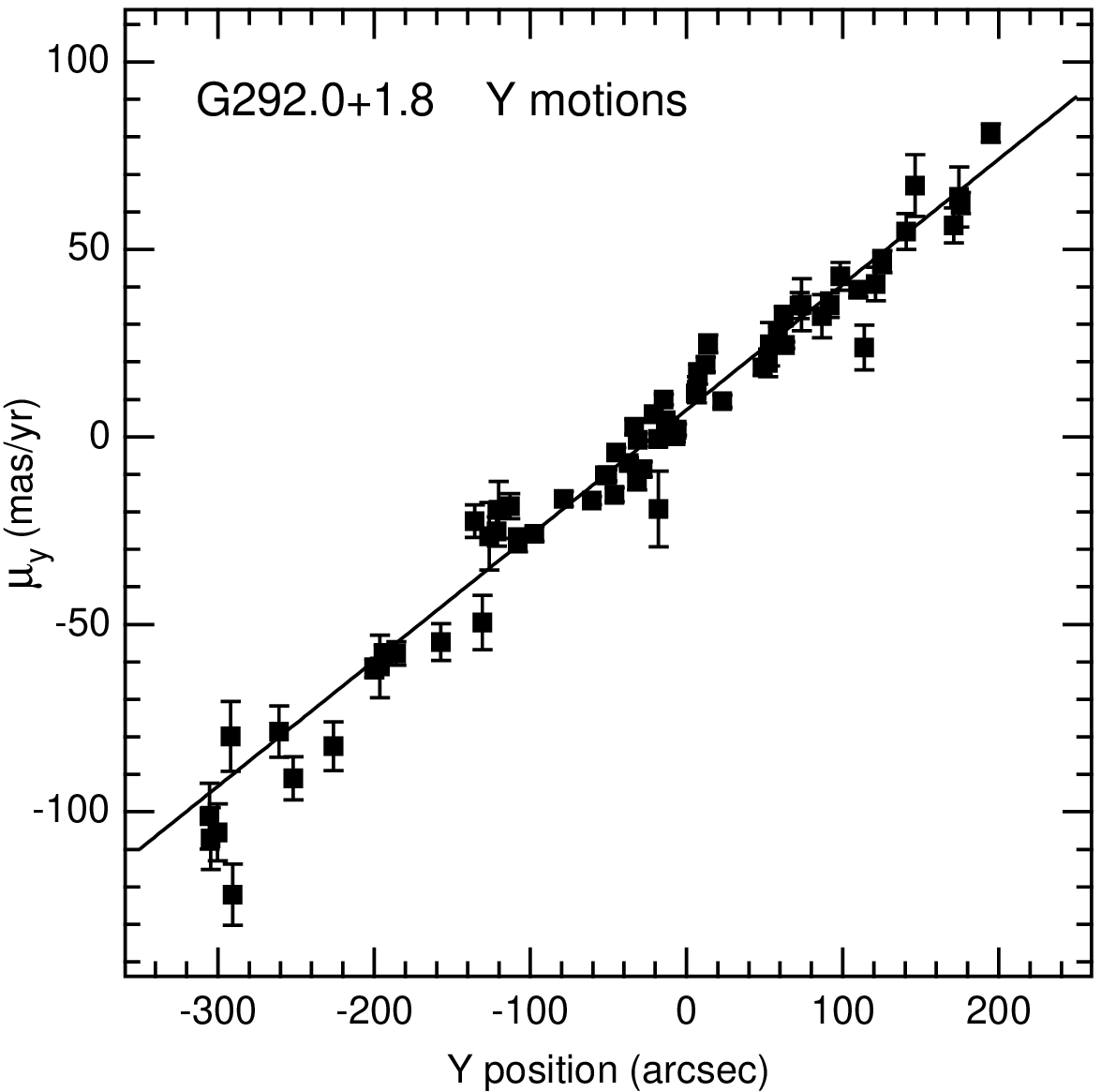}
\caption{Plots (in {\it x} and {\it y}) of the measured proper motions for 67 filaments in G292, as a function of position.  (The position is measured in arcsec relative to an arbitrary zero point near the center of G292.)  The error bars represent our best estimate of the uncertainties in measurements for individual filaments, according to the multiple-epoch technique described in \S 3.2\@.   The data are in good overall agreement  with a constant-velocity expansion model, represented by the best-fit lines.  The inverse slope of the line, constrained to be the same for both the {\it x-} and {\it y-}component fits, gives the age, and the intercepts with $\mu = 0$\ give the coordinates of the expansion center.}

\end{figure}

\clearpage

\begin{figure}
\epsscale{1.0}
\plotone{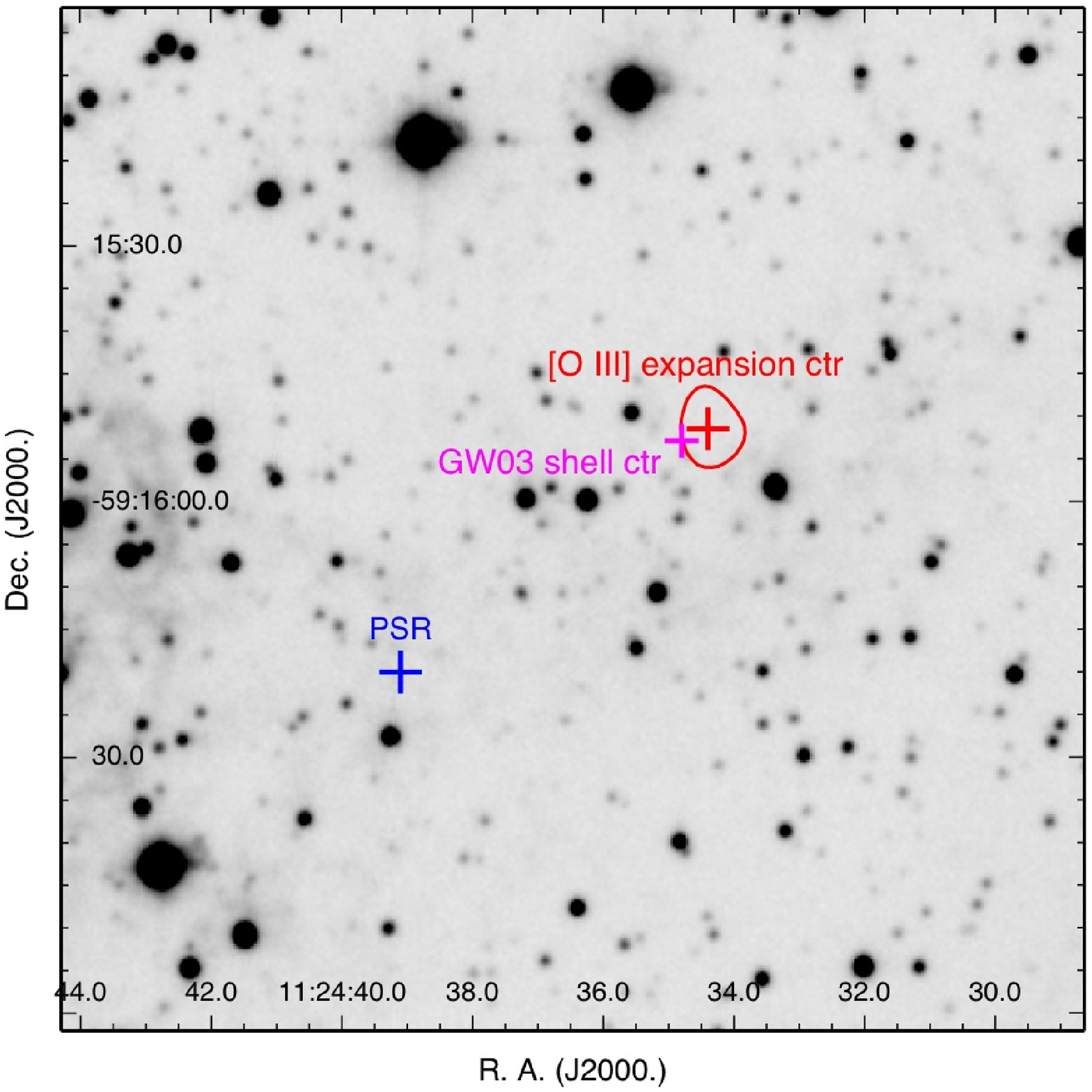}
\caption{In this 2\arcmin\ square central section of an unsubtracted \oiii \lam 5007 image of G292, our expansion center is indicated by the red cross, with the 90\%-confidence region also indicated in red.  The blue cross indicates the position of PSR J1124--5916, 46\arcsec\ SE of the expansion center.  The center of the outer radio shell, as given by \citet{gaensler03}, indicated by the magenta cross, is only 3\arcsec\ from our expansion center.  Some of the brighter \oiii\ filaments can be seen at the left (east) side of the figure.}
\end{figure}

\begin{figure}
\epsscale{0.95}
\plotone{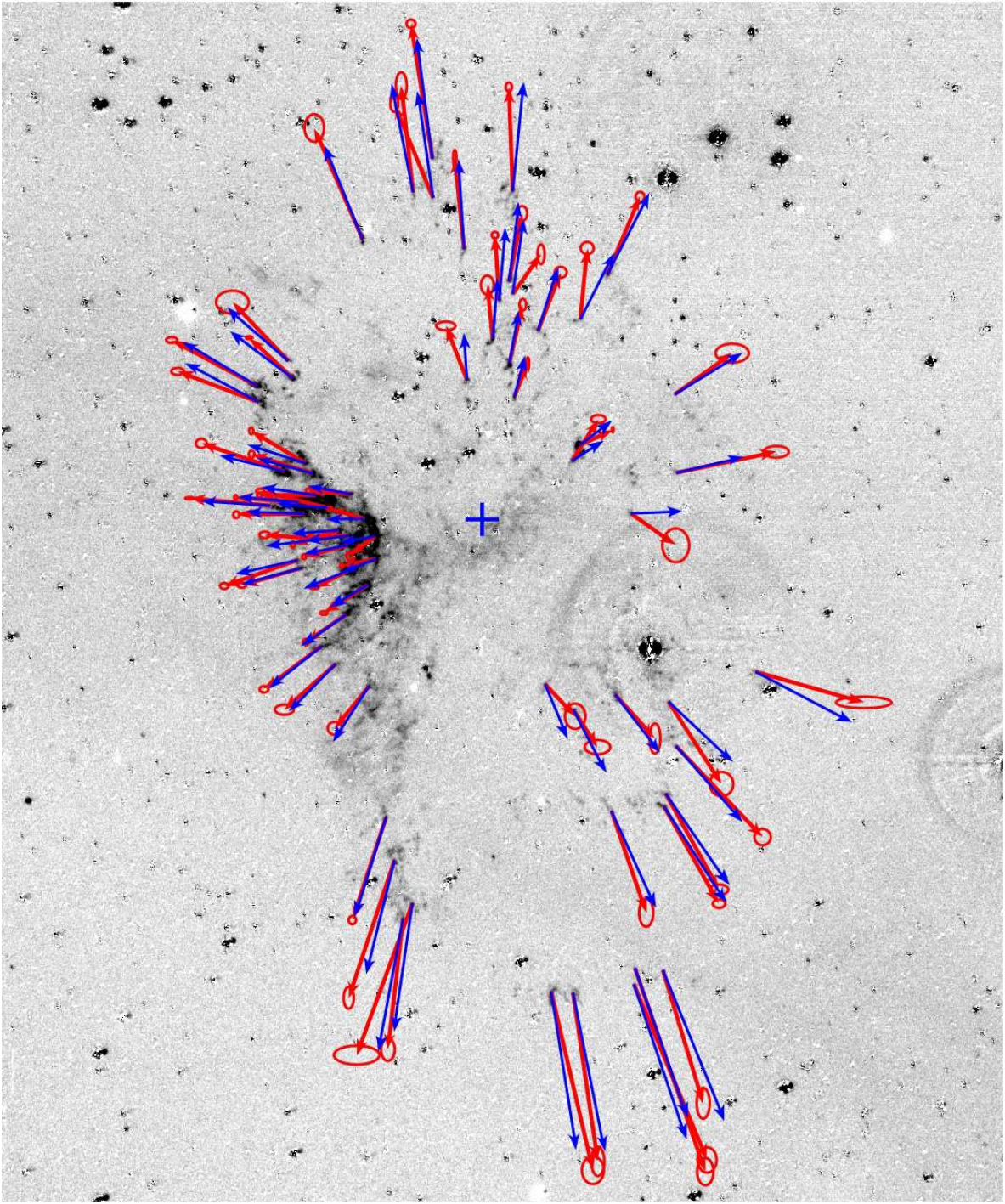}
\caption{The measured proper motions projected ahead 1000 yr are again represented as red vectors on the image of G292, exactly as in  Fig.~4, and in addition the 1000-yr proper motions for the best-fit model are represented as blue vectors.  Uncertainties in the measured values are represented by ellipses at the ends of the measured-value vectors.   The cross represents the best-fit expansion center, identical to that in Figs. 4 and 7\@.  While many of the model vectors lie well outside the corresponding error ellipse, there seems to be no strong pattern to the deviations.}
\end{figure}

\clearpage

\end{document}